# Cherenkov Friction and Radiation of a Neutral Polarizable Particle Moving near a Transparent Dielectric Plate

G. V. Dedkov and A. A. Kyasov

Kabardino-Balkarian State University, Nalchik, Russia

E-mail: gv_dedkov@mail.ru

**Abstract.** We have obtained general expressions for the intensity of radiation, decelerating force, the rate of heating and acceleration of a small polarizable particle under the conditions of Cherenkov friction: at relativistic motion parallel to the surface of thick transparent dielectric plate. Comparison with the results of other authors is given.

Casimir friction or quantum friction is the effect of quantum electrodynamics caused by quantum fluctuations of the electromagnetic field at relative movement in a vacuum of parallel smooth surfaces of two bodies or when a small particle moves parallel to the surface of an extended body (plate) [1, 2]. The processes associated with quantum and "thermal" (at a finite temperature of bodies) friction have aroused great interest of researchers during recent decades (see [3—5] and references).

An important consequence of quantum friction is the possibility of electromagnetic radiation under the Cherenkov condition $V > c/n$, where $V$ is the particle velocity (hereinafter we will consider just this case), $c$ is the speed of light in vacuum, and $n$ the refractive index of an immovable body (transparent dielectric plate) [6—8].

According to the interpretation of [7], the physical cause of radiation is an anomalous Doppler effect, at which the frequency of electromagnetic quanta absorbed by a particle is negative in the rest frame of the particle. As a result, the particle is excited into a quantum state with a higher energy and emits a photon within the Cherenkov cone. Unlike works [7, 8], in which the intensity of radiation was calculated using another method, we will use the general results of our theory of fluctuation-electromagnetic interaction of a relativistic particle moving near a surface [9, 10]. This makes it possible better understanding of the electrodynamic essence of the problem and to find the important relations between electrodynamical, mechanical and termal quantities. Our final expression for the intensity of radiation coincides with [7], but it differs somewhat from [8]. In contrast to [7], there are some differences in the relationship between radiation force, radiation power and the rate of change of the rest mass in the reference frame @ of an immovable plate.

We use the geometric configuration shown in the figure. The nature of the basic electrodynamic relationships with these quantities is determined by equations

$$\int \langle \mathbf{j} \cdot \mathbf{E} \rangle d^3 r = \langle \mathbf{\dot{d}} \cdot \mathbf{E} + \mathbf{\dot{m}} \cdot \mathbf{B} \rangle + \mathbf{V} \cdot \langle \nabla(\mathbf{d} \cdot \mathbf{E} + \mathbf{m} \cdot \mathbf{B}) \rangle = \dot{Q} + F_x V \ , \tag{1}$$

$$\dot{Q}' = \int_{\Omega'} \langle \mathbf{j}' \mathbf{E}' \rangle d^3 r' = \gamma^2 \langle \mathbf{\dot{d}} \cdot \mathbf{E} + \mathbf{\dot{m}} \cdot \mathbf{B} \rangle = \gamma^2 \dot{Q} \tag{2}$$

$$-\frac{dW}{dt} = \oint_\sigma \mathbf{S} \cdot d\vec{\sigma} + \int_\Omega \langle \mathbf{j} \cdot \mathbf{E} \rangle d^3 r \tag{3}$$

where $\gamma$ is the Lorentz-factor, $W = (1/8\pi)\int_\Omega (\langle \mathbf{E}^2 \rangle + \langle \mathbf{B}^2 \rangle) d^3 r$ is the energy of the field inside the volume $\Omega$ that is restricted by the wave surface surrounding the particle (see figure), $\mathbf{S} = (c/4\pi)\langle \mathbf{E} \times \mathbf{H} \rangle$ is the Pointing vector, $\mathbf{d}$ and $\mathbf{m}$ are the fluctuation electric and magnetic dipole moments of the particle, $\mathbf{E}, \mathbf{B}, \mathbf{H}$ and $\mathbf{j}$ are the fluctuation-electromagnetic fields and the electric current density; angular brackets and points above dipole moments denote complete quantum-statistical averaging and time differentiation. All unprimed variables refer to the frame of reference $\Sigma$. In formula (2) $\dot{Q}' = dQ'/dt'$ is the heating rate of a particle in its frame of reference $\Sigma'$. It should be noted that the quantity $\dot{Q} = dQ/dt$ has an independent meaning and coincides with the heating rate of the particle or surface only in the nonrelativistic limit. In the quasistationary case $dW/dt = 0$ from (3) it follows that

$$I = \oint_\sigma \mathbf{S} \cdot d\vec{\sigma} = -\int_\Omega \langle \mathbf{j} \cdot \mathbf{E} \rangle d^3 r \tag{4}$$

Moreover, from (1) and (4) it follows

$$I = -(dQ/dt + F_x V) \tag{5}$$

Formulas (1)—(5) were previously used in our calculations of the radiation of moving and rotating particles in a vacuum [11—13]. In the used configuration when the plate is transparent, all the above arguments also remain valid, and one can use the general results for $F_x$ and $\dot{Q}$ presented in [9, 10]

$$F_x = -\frac{\hbar\gamma}{2\pi^2}\int_0^\infty d\omega \int_{-\infty}^{+\infty} dk_x \int_{-\infty}^{+\infty} dk_y k_x \sum_{i=e,m} \alpha_i''(\gamma\omega^+)\,\mathrm{Im}\!\left(\frac{\exp(-2q_0 z)}{q_0} R_i(\omega,\mathbf{k})\right)$$
$$\cdot\left[\coth\!\left(\frac{\hbar\omega}{2k_B T_2}\right) - \coth\!\left(\frac{\gamma\hbar\omega^+}{2k_B T_1}\right)\right] + (\ldots) \tag{6}$$

$$\frac{dQ}{dt} = \frac{\hbar\gamma}{2\pi^2}\int_0^\infty d\omega \int_{-\infty}^{+\infty} dk_x \int_{-\infty}^{+\infty} dk_y \,\omega^+ \sum_{i=e,m} \alpha_i''(\gamma\omega^+)\,\mathrm{Im}\!\left(\frac{\exp(-2q_0 z)}{q_0} R_i(\omega,\mathbf{k})\right)$$
$$\cdot\left[\coth\!\left(\frac{\hbar\omega}{2k_B T_2}\right) - \coth\!\left(\frac{\gamma\hbar\omega^+}{2k_B T_1}\right)\right] + (\ldots) \tag{7}$$

where $\omega^+ = \omega + k_x V$, $T_1$ and $T_2$ are the local temperatures of particle (in $\Sigma'$) and the plate (in $\Sigma$), $\alpha_{e,m}''(\omega)$ are the imaginary components of the electric and magnetic polarizability, the terms (…) describe the interaction with vacuum modes of the electromagnetic field in the absence of plate, and make no contribution to further results. The auxiliary quantities in (6) and (7) are given by the expressions

$$R_e(\omega,\mathbf{k}) = \Delta_e(\omega)\!\left[2(k^2 - k_x^2\beta^2)(1-\omega^2/k^2 c^2) + (\omega^+)^2/c^2\right] +$$
$$+ \Delta_m(\omega)\!\left[2k_y^2\beta^2(1-\omega^2/k^2 c^2) + (\omega^+)^2/c^2\right] \tag{8}$$

$$R_m(\omega,\mathbf{k}) = \Delta_m(\omega)\!\left[2(k^2 - k_x^2\beta^2)(1-\omega^2/k^2 c^2) + (\omega^+)^2/c^2\right] +$$
$$+ \Delta_e(\omega)\!\left[2k_y^2\beta^2(1-\omega^2/k^2 c^2) + (\omega^+)^2/c^2\right] \tag{9}$$

$$\Delta_e(\omega) = \frac{q_0\varepsilon(\omega)-q}{q_0\varepsilon(\omega)+q},\quad \Delta_m(\omega) = \frac{q_0\mu(\omega)-q}{q_0\mu(\omega)+q},\quad q = \left(k^2 - (\omega^2/c^2)\varepsilon(\omega)\mu(\omega)\right)^{1/2},$$
$$q_0 = (k^2 - \omega^2/c^2)^{1/2},\quad k^2 = k_x^2 + k_y^2 \tag{10}$$

In the case of transparent dielectric $\varepsilon(\omega) = n^2$, $\mathrm{Im}\,\varepsilon(\omega) = 0$, $\mu(\omega) = 1$, and coefficients $\Delta_{e,m}(\omega)$ take the form

$$\Delta_e(\omega) = \frac{n^2\sqrt{k^2 - \omega^2/c^2} - \sqrt{k^2 - n^2\omega^2/c^2}}{n^2\sqrt{k^2 - \omega^2/c^2} + \sqrt{k^2 - n^2\omega^2/c^2}}, \quad \Delta_m(\omega) = \frac{\sqrt{k^2 - \omega^2/c^2} - \sqrt{k^2 - n^2\omega^2/c^2}}{\sqrt{k^2 - \omega^2/c^2} + \sqrt{k^2 - n^2\omega^2/c^2}} \quad (11)$$

Making use the limiting transitions $T_1 \to 0, T_2 \to 0$ in (6), (7) and taking into account the relations $\coth(\hbar\omega/2k_B T_2) \to sign(\omega)$, $\coth(\hbar\omega^+/2k_B T_1) \to sign(\omega^+)$ one obtains

$$F_x = \frac{2\hbar\gamma}{\pi^2} \int_0^\infty d\omega\, \theta(n\beta - 1) \int_{\omega/V}^{n\omega/c} dk_x k_x \int_0^{\sqrt{n^2\omega^2/c^2 - k_x^2}} dk_y \cdot \sum_{i=e,m} \alpha_i''(\gamma\omega^-) \operatorname{Im}\left(\frac{\exp(-2q_0 z_0)}{q_0} R_i(\omega, -k_x)\right) \quad (12)$$

$$\dot{Q} = \frac{2\hbar\gamma}{\pi^2} \int_0^\infty d\omega\, \theta(n\beta - 1) \int_{\omega/V}^{n\omega/c} dk_x \int_0^{\sqrt{n^2\omega^2/c^2 - k_x^2}} dk_y\, \omega^- \sum_{i=e,m} \alpha_i''(\gamma\omega^-) \operatorname{Im}\left(\frac{\exp(-2q_0 z_0)}{q_0} R_i(\omega, -k_x)\right) \quad (13)$$

where $R_i(\omega, -k_x)$ coincides with $R_{e,m}(\omega, \mathbf{k})$ at $\mathbf{k} = (-k_x, k_y)$ and $\omega^- = \omega - k_x V$.

Using (5), (12) and (13) we find

$$I = -\frac{2\hbar\gamma}{\pi^2} \int_0^\infty d\omega\, \theta(n\beta - 1) \int_{\omega/V}^{n\omega/c} dk_x \int_0^{\sqrt{n^2\omega^2/c^2 - k_x^2}} dk_y\, \omega \sum_{i=e,m} \alpha_i''(\gamma\omega^-) \operatorname{Im}\left(\frac{\exp(-2q_0 z_0)}{q_0} R_i(\omega, -k_x)\right) \quad (14)$$

where $\theta(x)$ – is the Heaviside step-function. Formulas (12) and (14) coincide with the analogous formulas in [7] with the difference that (12) and (14) also include the contribution of magnetic polarizability $\alpha_m''(\omega)$ of the particle. The limits of integration over the wave vectors in (12)—(14) correspond to the condition of the anomalous Doppler effect, since in the rest frame of particle $\Sigma'$ the photon frequency is negative: $\omega' = \gamma(\omega - k_x V) = \gamma\omega^- < 0$. Due to analytical properties of functions $\alpha_{e,m}''(\omega)$ and $\exp(-2q_0 z_0)/q_0$, from (12)—(14) it follows that $F_x < 0$, $\dot{Q} > 0$ and $I > 0$. This means that the conversion of the kinetic energy of particle into radiation proves to be the dominant mechanism in the process of Cherenkov friction (due to (5)). In this case, the particle itself heats up (see (2)). It is worth noting that formulas (12)—(14) are also valid in the case of refractive index dispersion: $n = n(\omega)$, if we assume that integration by frequencies is performed in the domain of transparency of the dielectric plate, $\varepsilon''(\omega) = 0$.

Let us focus briefly on the difference from the results in [7] as regards formula (13) and the equation of particle dynamics. With allowance for (2), the change in the particle rest mass caused by radiation can be written in the form

$$\frac{dm}{dt'} = \frac{\dot{Q}'}{c^2} = \gamma^2 \frac{\dot{Q}}{c^2} \qquad (15)$$

Taking into account $dt = \gamma\, dt'$, from (15) it follows

$$\frac{dm}{dt} = \gamma \frac{\dot{Q}}{c^2} \qquad (16)$$

We compare this result with equation (16) in [7], which in our designations has the form

$$-V \cdot F_x = I + \frac{dm}{\gamma\, dt'} \qquad (17)$$

From (17) it then follows

$$\frac{dm}{dt} = -(V \cdot F_x + I) \qquad (18)$$

In contrast to (18), formula (13) in [7] for $dm/dt$ contains an additional $\gamma$-factor in the right hand side. In this case, according to [7], $dm/dt = u^\mu F_\mu$, where $u^\mu$ and $F_\mu$ are the components of four-vectors of velocity and force, and from (18), (17) and (5) it follows (cf. with (16))

$$\frac{dm}{dt} = \frac{\dot{Q}}{c^2} \qquad (19)$$

The observed difference becomes significant at $\gamma \gg 1$, i. e. precisely when the intensity of Cherenkov radiation is maximal and may affect the solution of the dynamics equation,

$$\frac{d}{dt}\left(\frac{mc\beta}{\sqrt{1-\beta^2}}\right) = F_x \qquad (20)$$

which with allowance for (16) reduces to the form [12]

$$\gamma^3 mc \frac{d\beta}{dt} = F_x - \beta \gamma^2 \frac{\dot{Q}}{c} \qquad (21)$$

The right side of (21) can be simplified with allowance for the relation between the forces $F_x$ and $F'_x$ defined in the frames of reference $\Sigma$ and $\Sigma'$. The corresponding relation is obtained upon the time differentiation of the Lorentz transformation for the particle momentum

$$F'_x = F_x - \gamma V \frac{dm}{dt} \qquad (22)$$

From (16), (21) and (22) it follows that

$$F'_x = F_x - \beta\gamma^2 \frac{\dot{Q}}{c} \qquad (23)$$

$$mc\gamma^3 \frac{d\beta}{dt} = F'_x \qquad (24)$$

and the explicit expression for the force $F'_x$ is obtained when substituting (12) and (13) into (23)

$$F'_x = \frac{2\hbar\gamma^2}{\pi^2}\int_0^\infty d\omega\,\theta(n\beta-1)\int_{\omega/V}^{n\omega/c} dk_x (k_x - \beta\omega/c)\int_0^{\sqrt{n^2\omega^2/c^2 - k_x^2}} dk_y \sum_{i=e,m}\alpha_i''(\gamma\omega^-)\,\mathrm{Im}\left(\frac{\exp(-2q_0 z_0)}{q_0}R_i(\omega,-k_x)\right) \qquad (25)$$

In contrast to this, if we substitute Eq. (19) (instead (16)) into dynamics equation (20), then the formula for $F'_x$ obtained with allowance for (12), (13), (19) and (20) will differ from (25) since the factor $(k_x - \beta\omega/c)$ within the integrand expression in (25) is replaced by $k_x(\gamma^{-1} + \beta^2) - \beta\omega/c$.

In conclusion, it is worth noting that formulas (12) and (14) for the Cherenkov friction force and radiation power of a particle fully agree with results [7] obtained in an independent way, but the four-vector of force introduced in [7] leads to a discrepancy with the equation of dynamics, which follows from our analysis.

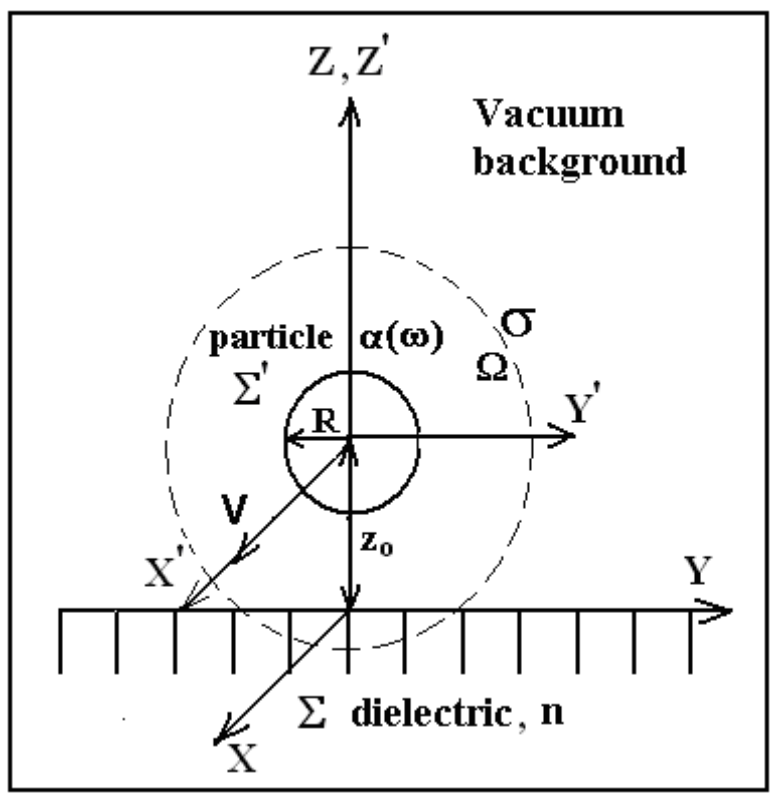

The geometric configuration and reference frames $\Sigma$ and $\Sigma'$ corresponding to the dielectric plate and particle. The dashed line shows the boundary of the wave zone of radiating particle.